\documentclass[10pt,conf]{IEEEtran}
\usepackage{amsmath,amssymb}
\usepackage{graphicx}
\usepackage{mathrsfs}
\usepackage{color}
\usepackage{tikz}
\usepackage{enumerate}
\usepackage{psfrag}	
\usepackage{array}
\usepackage{multirow,hhline}
\usepackage{exscale}
\usepackage{colortbl}
\usepackage{epsfig,subfigure}
\usepackage{cite}
\usepackage{amsthm}

\usepackage[outline]{contour}
\contourlength{1.2pt}
\usepackage{tikz}
\usepackage{pgfplots}
\usepackage{pgfplotstable}
\usepackage{rotate}
\usepgfplotslibrary{units}
\usetikzlibrary{%
patterns,%
calc,%
fit,%
arrows,%
plotmarks,%
shadows,%
chains,%
shapes%
}
\usepackage[outline]{contour}
\contourlength{1.2pt}
\tikzset{>=latex'} 
\tikzstyle{every picture}+=[remember picture] 
\definecolor{rubgray}{cmyk}{0.03,0.03,0.03,0.1}

\newtheorem{theorem}{Theorem}

\newcommand{\mat}[1]{\ensuremath{\boldsymbol{#1}}}
\renewcommand{\vec}[1]{\ensuremath{\boldsymbol{#1}}}
\newcommand{\y}[0]{\ensuremath{\boldsymbol{y}}}
\newcommand{\p}[0]{\ensuremath{\boldsymbol{p}}}
\newcommand{\z}[0]{\ensuremath{\boldsymbol{z}}}
\renewcommand{\H}[0]{\ensuremath{\boldsymbol{H}}}
\newcommand{\x}[0]{\ensuremath{\boldsymbol{x}}}
\newcommand{\X}[0]{\ensuremath{\boldsymbol{X}}}

\renewcommand{\c}[0]{\ensuremath{\boldsymbol{c}}}

\renewcommand{\a}[0]{\ensuremath{\boldsymbol{a}}}
\newcommand{\I}[0]{\ensuremath{\boldsymbol{I}}}
\newcommand{\D}[0]{\ensuremath{\boldsymbol{D}}}

\newcommand{\V}[0]{\ensuremath{\boldsymbol{V}}}
\newcommand{\U}[0]{\ensuremath{\boldsymbol{U}}}

\renewcommand{\d}[0]{\ensuremath{\boldsymbol{d}}}

\DeclareMathOperator{\tr}{trace}

\definecolor{Ogreen}{rgb}{.2,.4,.2}

\graphicspath{{./fig/}}

\begin{document}

\IEEEoverridecommandlockouts
\title{On Channel Inseparability and the DoF Region of MIMO Multi-way Relay Channels}
\author{
\IEEEauthorblockN{Anas Chaaban and Aydin Sezgin}\\
\IEEEauthorblockA{Institute of Digital Communication Systems\\
Ruhr-Universit\"at Bochum (RUB), Germany\\
Email: \{anas.chaaban, aydin.sezgin\}@rub.de
\vspace{-.5cm}}
\thanks{%
This work is supported by the German Research Foundation, Deutsche
Forschungsgemeinschaft (DFG), Germany, under grant SE 1697/5.
}
}

\maketitle

\thispagestyle{empty}

\begin{abstract}
Full-duplex multi-way relaying is a potential solution for supporting high data rates in future Internet-of-Things (IoT) and 5G networks. Thus, in this paper the full-duplex MIMO multi-way channel consisting of 3 users (Y-channel) with $M$ antennas each and a common relay node with $N$ antennas is studied. Each user wants to exchange messages with all the other users via the relay. A transmission strategy is proposed based on channel diagonalization that decomposes the channel into parallel sub-channels, and physical-layer network coding over these sub-channels. It is shown that the proposed strategy achieves the optimal DoF region of the channel if $N\leq M$. Furthermore, the proposed strategy that requires joint encoding over multiple sub-channels is compared to another strategy that encodes over each sub-channel separately. It turns out that coding jointly over sub-channels is necessary for an optimal transmission strategy. This shows that the MIMO Y-channel is inseparable.
\end{abstract}


\section{Introduction}
It is predicted that the number of devices with communication capability will rise to 50 billions in the upcoming years \cite{Evans_IoT}. The resulting web of devices connected by the Internet-of-Things (IoT) and Machine-to-Machine (M2M) communications will lead to more sophisticated network topologies, where relaying will play a key role in enabling reliable communications. One integral relaying capability which will be of great use in the future is multi-way relaying.

Multi-way relaying refers to situations where multiple users want to exchange information via a common relay node. The multi-way relay channel (MWRC) with 2-users is known as the two-way relay channel TWRC. The TWRC has been studied thoroughly recently \cite{KimDevroyeMitranTarokh, AvestimehrSezginTse, ShaqfehZafarAlnuweiriAlouini}. Although the TWRC has become well-understood recently (capacity characterization within a constant gap), the multiple user case is still not completely understood. Partial characterizations for the multiple-user MWRC are obtained in \cite{ChaabanSezginAvestimehr_YC_SC, MatthiesenZapponeJorswieck, GunduzYenerGoldsmithPoor_IT, MokhtarMohassebNafieElGamal, SezginAvestimehrKhajehnejadHassibi}. In this paper, we are interested in the multi-way relay channel with 3-users known as the Y-channel. The users of the Y-channel exchange information in all directions via the relay. The extension of the TWRC to the Y-channel is not straightforward, and many challenges have to be tackled when making this step. One of the challenges is in deriving capacity upper bounds. While the capacity of the TWRC can be approximated with high-precision using the cut-set bounds \cite{CoverThomas}, the capacity of the Y-channel requires new bounds. Such bounds have been derived in \cite{ChaabanSezginAvestimehr_YC_SC}. Another challenge is in the communication strategy. The communication strategies of the TWRC do not suffice for approaching the capacity of the Y-channel and new strategies have to be developed \cite{ChaabanSezgin_ISIT12_Y, ZewailNafieMohassebGamal}. 


In this paper, we address the degrees-of-freedom (DoF) region of the MIMO Y-channel \cite{LeeLimChun}. While the sum-DoF is known \cite{LeeLimChun,ChaabanOchsSezgin}, a complete DoF region characterization is not available to-date, except for some classes of MIMO Y-channels \cite{ZewailNafieMohassebGamal}. The importance of the DoF region is that it reflects the trade-off between the achievable DoF of different users, contrary to the sum-DoF which does not.

We develop a communication strategy for the MIMO Y-channel with $M$ antennas at the users, and $N\leq M$ antennas at the relay. Our proposed strategy revolves around two ideas: channel diagonalization, and physical-layer network-coding. Channel diagonalization is performed by zero-forcing beam-forming. After channel diagonalization, the MIMO Y-channel is decomposed into a set of parallel SISO Y-channels (sub-channels). Then, bi-directional, cyclic, and uni-directional communication is performed over these sub-channels. Bi-directional communication ensures the exchange of information between two users as in the TWRC. Cyclic communication ensures information exchange over users in a cyclic manner, such as exchanging a signal from user $i$ to $j$, $j$ to $k$, an $k$ to $i$. Both bi-directional and cyclic communication are based on compute-forward \cite{NazerGastpar}, while uni-directional communication is based on decode-forward. We also provide a resource allocation strategy for distributing the sub-channels among the users. Finally, we show that the optimized strategy achieves the DoF region of the channel with $N\leq M$. The converse is provided by a DoF region outer bound based on an upper bound that was derived in \cite{ChaabanOchsSezgin} for characterizing the sum-DoF of the Y-channel.\footnote{The main difference between this work and \cite{ChaabanOchsSezgin_EW} is the incorporation of cyclic communication. Cyclic communication combined with the transmission scheme in \cite{ChaabanOchsSezgin_EW} completes the achievability of the DoF region.} 

As a by-product of this work, it is concluded that the MIMO Y-channel is inseparable \cite{CadambeJafar_Inseperability}. That is, the parallel sub-channels of the MIMO Y-channel have to treated jointly, since separate encoding over each sub-channel is not optimal. 

Throughout the paper, we use bold-face lower-case and upper-case letters to denote vectors and matrices, respectively, and normal-face letters to denote scalars. The notation $\X^{H}$ and $\X^{-1}$ denotes the Hermitian transpose and the inverse of $\X$, respectively. We say that $\x\sim\mathcal{CN}(\vec{m},\mat{Q})$ if $\x$ is a complex Gaussian random vector with mean $\vec{m}$ and covariance matrix $\mat{Q}$. We use $\I_N$ to denote the $N\times N$ identity matrix and $\vec{0}_q$ to denote the $q\times 1$ zero vector.


\begin{figure*}[ht]
\centering
\subfigure[Uplink.]{
\includegraphics[width=.8\columnwidth]{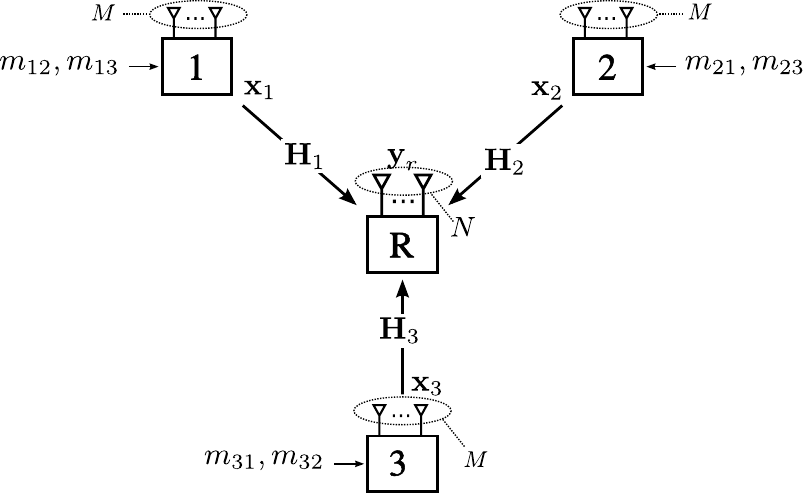}
%
%
%
%
\label{Fig:ModelU}
}
\hspace{1cm}
\subfigure[Downlink.]{
\includegraphics[width=.8\columnwidth]{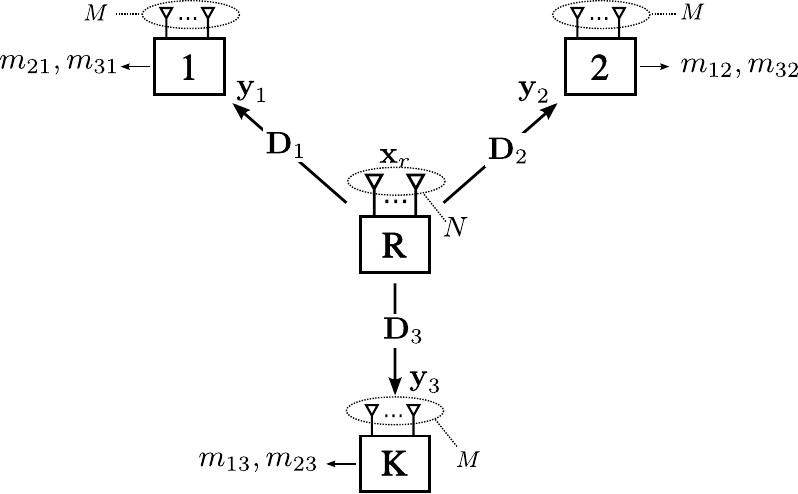}
\label{Fig:ModelD}
}
\caption{In the MIMO Y-channel, each user sends a message to each of the remaining users. Consequently, each user decodes 2 messages.}
\label{Fig:Ychannel}
\end{figure*}

\section{System Model}
\label{Sec:Model}
The MIMO Y-channel consists of 3 users which want to establish full message-exchange via a relay as shown in Figures \ref{Fig:ModelU} and \ref{Fig:ModelD}. All nodes are assumed to be full-duplex with power $\rho$. The relay has $N$ antennas, and each user has $M$ antennas. User $i\in\mathcal{K}=\{1,2,3\}$ has messages $m_{ij}$ and $m_{ik}$ to be sent to users $j\in\mathcal{K}\setminus\{i\}$ and $k\in\mathcal{K}\setminus\{i,j\}$, respectively. The rate of message $m_{ij}$ is $R_{ij}(\rho)$. 

At time instant $t$, user $i$ sends a signal $\x_{i}(t)\in\mathbb{C}^M$ which is a function of the messages $m_{ij}$ and $m_{ik}$, and the received signal up to that time instant $\y_i^{t-1}$. The received signal at the relay is given by (cf. Figure \ref{Fig:ModelU})
\begin{align}
\y_{r}(t)=\sum_{i=1}^3\H_i\x_i(t)+\z_{r}(t),
\end{align}
which is an $N\times1$ vector, where $\z_{r}(t)\sim\mathcal{CN}(\vec{0}_N,\vec{I}_N)$. Here $\H_i$ is the $N\times M$ complex channel matrix from user $i$ to the relay, which is assumed to be constant during the duration of the transmission. The relay transmit signal at time $t$ is denoted $\x_{r}(t)\in\mathbb{C}^N$ and is a function of the received signal at the relay up to that time instant $\y_r^{t-1}$. The received signal at user $i$ is given by (cf. Fig. \ref{Fig:ModelD})
\begin{align}
\label{ReceivedSignal}
\y_{i}(t)=\D_i\x_{r}(t)+\z_{i}(t),
\end{align}
which is an $M\times1$ vector, where $\z_{i}(t)\sim\mathcal{CN}(\vec{0}_M,\vec{I}_M)$, and $\D_i$ is the $M\times N$ downlink complex (static) channel matrix from the relay to user $i$. The transmit signals of the users and the relay must satisfy the power constraint $\rho$, i.e., $\tr(\mathbb{E}[\x_i\x_i^{H}])\leq \rho$ and $\tr(\mathbb{E}[\x_r\x_r^{H}])\leq \rho$.

The achievable rates and the capacity region of the MIMO Y-channel are defined in the standard information-theoretic sense~\cite{CoverThomas}. Since we are interested in the DoF region of the channel, we define the the DoF of message $m_{ij}$ as \cite{CadambeJafar_KUserIC}
\begin{align}
\label{DoFDef}
d_{ij}=\lim_{\substack{\rho\to\infty}}\frac{R_{ij}(\rho)}{\log(\rho)}.
\end{align}
A DoF $d_{ij}$ is said to be achievable if there exists an achievable rate $R_{ij}(\rho)$ satisfying \eqref{DoFDef}. A DoF tuple $\d\in\mathbb{R}^{6}$ defined as $\d=(d_{12},d_{13},d_{21},d_{23},d_{31},d_{32})$ is said to be achievable if all its components are simultaneously achievable. We define the DoF region $\mathcal{D}$ as the set of all achievable DoF tuples $\d$. We also define the sum-DoF as $d_\Sigma=\max_{\d\in\mathcal{D}}(d_{12}+d_{13}+d_{21}+d_{23}+d_{31}+d_{32})$. Now, we are ready to present the main result of the paper given in the next section.

\section{Main result}
\label{SecMainResult}
The main result of the paper is a characterization of the DoF region of the MIMO Y-channel with $N\leq M$, as given in the following theorem.
\begin{theorem}
\label{Thm:DoF}
The DoF region $\mathcal{D}$ of the MIMO Y-channel with $N\leq M$ is given by the set of tuples $\d\in\mathbb{R}^6$ satisfying 
\begin{align}
\label{Eq:DoFBoundThm}
\sum_{i=1}^{2}\sum_{j=i+1}^{3}d_{p_ip_j}\leq N, \quad \forall \mathbf{p}
\end{align}
where $\mathbf{p}$ is a permutation of $\mathcal{K}$ and $p_i$ is its $i$-th component.
\end{theorem}
The converse of Theorem \ref{Thm:DoF} is based on an upper bound derived in \cite{ChaabanOchsSezgin}. The achievability of this theorem is the main focus of the rest of the paper. The achievability is proved by using three steps: (i) channel diagonalization, (ii) physical-layer network coding, and (iii) resource allocation. Channel diagonalization is performed by zero-forcing pre- and post-coding, which transforms the channel into a set of parallel SISO Y-channels. Next, compute-forward \cite{NazerGastpar} and decode-forward \cite{CoverElgamal} strategies are applied over these sub-channels. Finally, a resource (sub-channel) allocation is performed that achieves the outer bound in Theorem \ref{Thm:DoF}. 

\begin{figure}
\centering
\begin{tikzpicture}[semithick,scale=.9, every node/.style={scale=.9}]
\node[circle] (c1) at (0,0) [fill=white, draw] {$p_1$};
\node[circle] (c2) at (3,0) [fill=white, draw] {$p_2$};
\node[circle] (c3) at (6,0) [fill=white, draw] {$p_3$};
\draw[->] (c1.east) to node[above] {$d_{p_1p_2}$} (c2.west) ;
\draw[->] (c2.east) to node[above] {$d_{p_2p_3}$}  (c3.west);
\draw[->] ($(c1.east)+(0,.2)$) to [out=45,in=135] node[above] {$d_{p_1p_3}$}  ($(c3.west)+(0,.2)$);
\end{tikzpicture}
\caption{Message flow graph for the 3-user Y-channel representing the DoF upper bound \eqref{Eq:DoFBoundThm}. Note that this graph has no cycles.}
\label{Fig:MessageFlow3Users}
\end{figure}
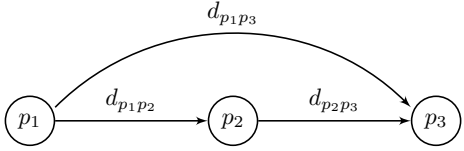

Before we proceed with the proof of achievability, we would like to highlight some properties of the upper bound in \eqref{Eq:DoFBoundThm}. Let us represent this upper bound by a message-flow graph as shown in Fig. \ref{Fig:MessageFlow3Users}. In this graph, each user is represented by a node, and each DoF component is represented by a directed edge from the source node to the destination node. Notice the following interesting properties of this graph:
\begin{itemize}
\item[(a)] It has only 3 edges.
\item[(b)] It has no cycles.
\end{itemize}
These properties are true for any permutation $\p$ of the users. This is a key observation for the design of the communication strategy that achieves this upper bound. We need to design strategies which do not lead to bounds which violate these properties. Next, we consider a simple toy-example to motivate our communication strategy.

\vspace{-.2cm}

\section{Toy example}
\label{Sec:3UserYChannel}
Consider the DoF tuple $\hat{\d}=(2,0,1,1,1,0)$ to be achieved over a Y-channel with $M=N=3$. According to Theorem \ref{Thm:DoF}, this DoF tuple is achievable since it belongs to $\mathcal{D}$. How can we achieve this DoF tuple? To answer this question, let us start by examining a uni-directional strategy over the Y-channel.

\vspace{-.2cm}

\subsection{A uni-directional strategy}
A uni-directional strategy is a simple communication strategy that ensures a uni-directional information flow over a relay channel, such as a decode-forward (DF) strategy \cite{CoverElgamal,ShaqfehQahtaniAlnuweiri}. In DF, the uplink and downlink can be modelled as MIMO multiple-access and broadcast channels, respectively. From signal-space dimensions point of view,  each symbol communicated using uni-directional strategy consumes one dimension at the relay. Thus, the achievability of $\hat{\d}$ would require that the sum-DoF satisfies
\begin{align}
\label{Eq:UniDirBound3User}
d_\Sigma=d_{12}+d_{13}+d_{21}+d_{23}+d_{31}+d_{32}\leq N,
\end{align}
which is not true for $\hat{\d}$ since it sums up to $d_\Sigma=5>N$. Thus, such a uni-directional strategy is not able to achieve $\hat{\d}$.

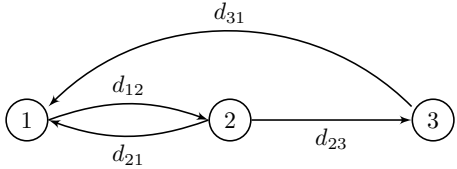
\begin{figure}
\centering
\begin{tikzpicture}[semithick,scale=.9, every node/.style={scale=.9}]
\node[circle] (c1) at (0,0) [fill=white, draw] {$1$};
\node[circle] (c2) at (3,0) [fill=white, draw] {$2$};
\node[circle] (c3) at (6,0) [fill=white, draw] {$3$};
\draw[->] (c1.east) to [out=20,in=160] node[above] {$d_{12}$} (c2.west) ;
\draw[->] (c2.west) to [out=200,in=-20] node[below] {$d_{21}$} (c1.east) ;
\draw[->] (c2.east) to node[below] {$d_{23}$}  (c3.west);
\draw[->] ($(c3.west)+(0,.2)$) to [out=135,in=45] node[above] {$d_{31}$}  ($(c1.east)+(0,.2)$);
\end{tikzpicture}
\caption{Message-flow graph corresponding to \eqref{Eq:UniDirBound3User}. Note the 2-cycle $(1,2)$ and the 3-cycle $(1,2,3)$.}
\label{Fig:MessageFlow3UsersDHat}
\end{figure}

Let us study the properties of the bound \eqref{Eq:UniDirBound3User} using the message-flow graph in Fig. \ref{Fig:MessageFlow3UsersDHat}. One can easily see that \eqref{Eq:UniDirBound3User} violates both properties (a) and (b) of the upper bound \eqref{Eq:DoFBoundThm} since it has 4 edges and it also has the 2-cycle $(1,2)$, and the 3-cycle $(1,2,3)$.\footnote{An $\ell$-cycle (cycle of length $\ell$) is denoted by a tuple $\c=(v_1,\cdots,v_\ell)$ with corresponding edges $(v_1,v_2), (v_2,v_3),\cdots,(v_\ell,v_1)$. Note that $\c$ is cyclic-shift invariant, i.e., the $\ell$-cycle $(v_\ell,v_1,\cdots,v_{\ell-1})$ is equivalent to $\c$.} To achieve $\hat{\d}$, we need to resolve these violations. Let us first deal with 2-cycles.

\subsection{A bi-directional strategy}
\label{Sec:BiDirStra3User}
The bi-directional strategy that is commonly used in the TWRC \cite{NamChungLee, AvestimehrSezginTse} resolves 2-cycles. In our particular example, let user 1 and 2 send symbols $u_{12}$ and $u_{21}$, respectively, along one dimension at the relay. In this case, the relay can compute a linear combination $L(u_{12},u_{21})$ of these symbols, and forward this to users 1 and 2 in the downlink over one dimension. Then, each user can decode the desired signal after subtracting his own self-interference. This operation requires 1 dimension to send 2 symbols leading to an efficiency of 2 DoF/dimension, which is better than the uni-directional strategy with 1 DoF/dimension. After this operation, it remains to achieve $\tilde{\d}=(1,0,0,1,1,0)$. If we were to achieve $\tilde{\d}$ using the uni-directional strategy, we would need 3 more dimensions. In total, the combination of bi-directional and uni-directional strategies would require $d_{12}+d_{23}+d_{31}$ to satisfy
\begin{align}
\label{Eq:ThCycle}
d_{12}+d_{23}+d_{31}\leq N.
\end{align}
But $d_{12}+d_{23}+d_{31}=4>N$ and thus $\hat{\d}$ is still not achievable by combining the bi-directional and uni-directional strategies (although this reduced the required dimensions from 5 to 4). Note that the message-flow graph corresponding to \eqref{Eq:ThCycle} fulfils property (a), but not property (b) since it contains a 3-cycle. This problem is resolved next.

\vspace{-.2cm}

\subsection{A cyclic strategy}
After assigning 1 dimension to the bi-directional strategy, 2 dimensions remain available at the relay for achieving $\tilde{\d}$. Let users 1 and 2 align symbols $v_{12}$ and $v_{23}$ along one dimension at the relay, and let users 2 and 3 align symbols $v_{23}$ and $v_{31}$ along another dimension at the relay, respectively. Here, $v_{23}$ is sent twice by user 2, each time along a different dimension. Since the relay has 2 dimensions remaining at its disposal, the relay can compute linear combinations $L_1(v_{12},v_{23})$ and $L_2(v_{23},v_{31})$, and forward these combinations in the downlink over 2 dimensions. After reception, user 1 subtracts the self-interference $v_{12}$ from $L_1$ and decodes $v_{23}$, and then subtracts $v_{23}$ from $L_2$ and decodes $v_{31}$. Similarly users 2 and 3 can obtain the unknown symbols.

\begin{figure*}
\centering
\begin{tikzpicture}[semithick,scale=.85, every node/.style={scale=.85}]
\node (t1) at (0,0) {};
\node (t2) at (0,-2.5) {};
\node (t3) at (0,-5) {};
\node (r) at (6,-2.5) {};
\node (r1) at (12,0) {};
\node (r2) at (12,-2.5) {};
\node (r3) at (12,-5) {};
\node[fill=white] at ($(t1)+(-1.0,1.2,0)$) {User 1};
\node[fill=white] at ($(t2)+(-1.0,1.2,0)$) {User 2};
\node[fill=white] at ($(t3)+(-1,1.2,0)$) {User 3};
\node[fill=white] at ($(r)+(-1,1.2,0)$) {Relay};
\node[fill=white] at ($(r1)+(-1,1.2,0)$) {User 1};
\node[fill=white] at ($(r2)+(-1,1.2,0)$) {User 2};
\node[fill=white] at ($(r3)+(-1,1.2,0)$) {User 3};

\draw [] ($(t1)-(1.6,.65,0)$) rectangle ($(t1)+(1.6,1.45)$);
\draw [] ($(t2)-(1.6,.65,0)$) rectangle ($(t2)+(1.6,1.45)$);
\draw [] ($(t3)-(1.6,.65,0)$) rectangle ($(t3)+(1.6,1.45)$);
\draw [] ($(r)-(1.6,.65,0)$) rectangle ($(r)+(1.6,1.45)$);
\draw [] ($(r1)-(1.6,.65,0)$) rectangle ($(r1)+(1.6,1.45)$);
\draw [] ($(r2)-(1.6,.65,0)$) rectangle ($(r2)+(1.6,1.45)$);
\draw [] ($(r3)-(1.6,.65,0)$) rectangle ($(r3)+(1.6,1.45)$);

\foreach \i in {t1,t2,t3,r,r1,r2,r3}
{ 
\draw[->,dotted] (\i.center) to ($(\i)+(1.4,0,0)$);
\draw[->,dotted] (\i.center) to ($(\i)+(0,1.4,0)$);
\draw[->,dotted] (\i.center) to ($(\i)+(0,0,1.5)$);
}

\draw[->,blue] (t1.center) to ($(t1)+(1.5,1.3,.5)$);
\node[blue] at ($(t1)+(0.7,1.1,0)$) {$u_{12}$};
\draw[->,blue] (t2.center) to ($(t2)+(1.2,1.2,0)$);
\node[blue] at ($(t2)+(0.7,1.1,0)$) {$u_{21}$};
\draw[->,blue] (r.center) to ($(r)+(.8,0.8,0)$);
\node[blue] at ($(r)+(.3,0.7,0)$) {$u_{12}$};
\draw[->,blue] (r.center) to ($(r)+(1.2,1.2,0)$);
\node[blue] at ($(r)+(.8,1.2,0)$) {$u_{21}$};
\draw[->,blue] (r1.center) to ($(r1)+(1,1,0)$);
\node[blue] at ($(r1)+(.8,1.1,0)$) {$u_{12}+u_{21}$};
\draw[->,blue] (r2.center) to ($(r2)+(1,1,0)$);
\node[blue] at ($(r2)+(.8,1.1,0)$) {$u_{12}+u_{21}$};
\draw[->,blue] (r3.center) to ($(r3)+(1,1,0)$);
\node[blue] at ($(r3)+(.8,1.1,0)$) {$u_{12}+u_{21}$};

\draw[->,red] (t1.center) to ($(t1)+(0,1.3,1.8)$);
\node[red] at ($(t1)+(.3,1.5,2)$) {$v_{12}$};
\draw[->,red] (t2.center) to ($(t2)+(0,1.1,1.1)$);
\node[red] at ($(t2)+(.3,1.5,1.7)$) {$v_{23}$};
\draw[->,red] (r.center) to ($(r)+(0,1.0,1.0)$);
\node[red] at ($(r)+(-.1,.7,1.2)$) {$v_{12}$};
\draw[->,red] (r.center) to ($(r)+(0,1.5,1.5)$);
\node[red] at ($(r)+(-.1,1.5,2)$) {$v_{23}$};
\draw[->,red] (r1.center) to ($(r1)+(0,.9,.9)$);
\node[red] at ($(r1)+(-.3,1.3,1.4)$) {$v_{12}+v_{23}$};
\draw[->,red] (r2.center) to ($(r2)+(0,.9,.9)$);
\node[red] at ($(r2)+(-.3,1.3,1.4)$) {$v_{12}+v_{23}$};
\draw[->,red] (r3.center) to ($(r3)+(0,.9,.9)$);
\node[red] at ($(r3)+(-.3,1.3,1.4)$) {$v_{12}+v_{23}$};

\draw[->,Ogreen] (t2.center) to ($(t2)+(1.5,0,1.0)$);
\node[Ogreen] at ($(t2)+(0.6,0,1.0)$) {$v_{23}$};
\draw[->,Ogreen] (t3.center) to ($(t3)+(1,1,0)$);
\node[Ogreen] at ($(t3)+(.5,.9,0)$) {$v_{31}$};
\draw[->,Ogreen] (r.center) to ($(r)+(.8,0,0.6)$);
\node[Ogreen] at ($(r)+(0.5,0,.8)$) {$v_{31}$};
\draw[->,Ogreen] (r.center) to ($(r)+(1.2,0,0.9)$);
\node[Ogreen] at ($(r)+(1.2,0,1.3)$) {$v_{23}$};
\draw[->,Ogreen] (r1.center) to ($(r1)+(.8,0,.8)$);
\node[Ogreen] at ($(r1)+(1.3,0.05,1.3)$) {$v_{23}+v_{31}$};
\draw[->,Ogreen] (r2.center) to ($(r2)+(.8,0,.8)$);
\node[Ogreen] at ($(r2)+(1.3,0.05,1.3)$) {$v_{23}+v_{31}$};
\draw[->,Ogreen] (r3.center) to ($(r3)+(.8,0,.8)$);
\node[Ogreen] at ($(r3)+(1.3,0.05,1.3)$) {$v_{23}+v_{31}$};

\draw[->] ($(t1)+(2.0,0.4,0)$) to node[] {\contour{white}{$\H_{1}$}} ($(r)-(1.8,-1.1,0)$) ;
\draw[->] ($(t2)+(2.0,0.4,0)$) to node[] {\contour{white}{$\H_{2}$}} ($(r)-(1.8,-0.4,0)$) ;
\draw[->] ($(t3)+(2.0,.4,0)$) to node[] {\contour{white}{$\H_{3}$}} ($(r)-(1.8,0.5,0)$) ;
\draw[->] ($(r)+(1.9,1.1,0)$) to node[] {\contour{white}{$\D_{1}$}} ($(r1)-(1.8,-0.4,0)$) ;
\draw[->] ($(r)+(1.9,.4,0)$) to node[] {\contour{white}{$\D_{2}$}} ($(r2)-(1.8,-0.4,0)$) ;
\draw[->] ($(r)+(1.9,-.5,0)$) to node[] {\contour{white}{$\D_{3}$}} ($(r3)-(1.8,-0.4,0)$) ;

\end{tikzpicture}
\caption{A graphical illustration of the transmitter, relay, and receiver signal-space for the toy-example in Section \ref{Sec:3UserYChannel}. The relay computes the sum of the symbols received along each of the three directions, and forwards these sums. Each user is able to extract his desires signals after subtracting self-interference.}
\label{Fig:SigSpace3User}
\end{figure*}
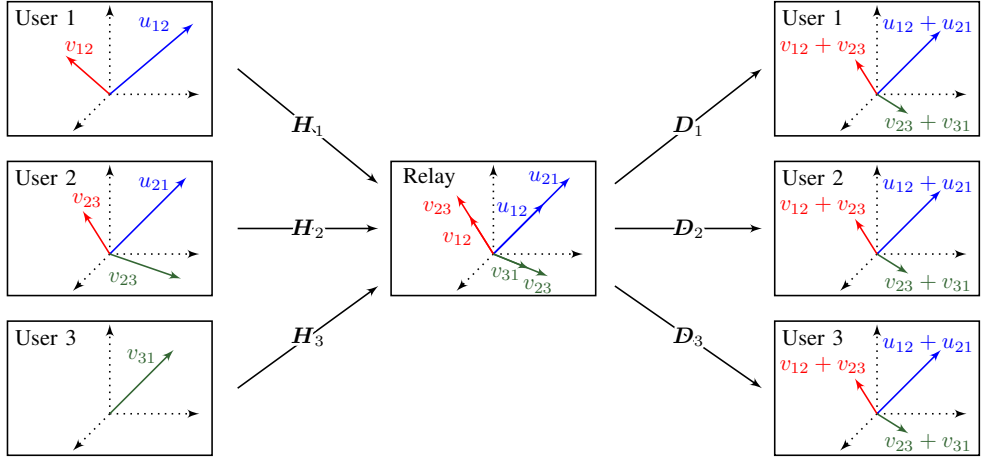

This strategy requires only two dimensions at the relay, contrary to the uni-directional strategy which requires 3 dimensions to deliver the same symbols. After this step, the DoF tuple $\hat{\d}$ is achieved. The resulting user and relay signal-space is as shown in Figure \ref{Fig:SigSpace3User}. Note that the uni-directional strategy was not required in the final scheme in this particular toy-example. This is not true in general as shown next.

\vspace{-.2cm}

\section{Achievability of Theorem \ref{Thm:DoF}}
\label{Sec:Achievability}
Now we are ready to prove the achievability of Theorem \ref{Thm:DoF}. In this section, we will describe a transmission scheme based on channel diagonalization and a combination of bi-directional, cyclic, and uni-directional transmission strategies, and we will propose an optimal resource allocation strategy for this network. We start with channel diagonalization.

\subsection{Channel diagonalization}
Channel diagonalization is performed by using zero-forcing beam-forming with the aid of the Moore-Penrose pseudo inverse (MPPI) to diagonalize the uplink and the downlink channels simultaneously. To this end, the transmit signal of user $i$ is constructed as $\x_i=\V_i\a_i$ where $\a_i\in\mathbb{C}^{N\times 1}$ is a vector which contains the codeword symbols satisfying $\tr(\mathbb{E}[\a_i\a_i^{H}])=\rho$, $\V_i$ is a pre-coding matrix given by $$\V_i=\alpha_i\H_i^{H}[\H_i\H_i^{H}]^{-1},$$ and $\alpha_i$ is a normalization coefficient that guarantees $\tr(\mathbb{E}[\x_i\x_i^{H}])=\rho$. The matrix $\V_i$ exists if $N\leq M$. Using this construction, the relay received signal is given by
\begin{align}
\y_r&=\alpha_1\a_1+\alpha_2\a_2+\alpha_3\a_3+\z_r.
\end{align}
This achieves channel diagonalization in the uplink. For sub-channel $s=1,\cdots,N$, we get
\begin{align}
y_{r,s}=\alpha_1 a_{1,s}+\alpha_2 a_{2,s}+\alpha_3 a_{3,s}+z_{r,s},
\end{align}
where $y_{r,s}$, $a_{i,s}$, and $z_{r,s}$ are the $s$-th components of $\y_r$, $\a_i$, and $\z_r$, respectively. In the downlink, user $i$ post-codes the received signal using the post-coding matrix $$\U_i=[\D_i^{H}\D_i]^{-1}\D_i^{H},$$ which exists if $N\leq M$. The post-coded signal is given by
\begin{align}
\tilde{\y}_i=\U_i\y_i=\x_r+\tilde{\z}_i,
\end{align}
where $\tilde{\z}_i$ is the processed noise at user $i$. This achieves channel diagonalization in the downlink, and user $i$ gets 
\begin{align}
\tilde{y}_{i,s}=x_{r,s}+\tilde{z}_{i,s},
\end{align}
over the $s$-th sub-channel, where $\tilde{y}_{i,s}$, $x_{r,s}$, and $\tilde{z}_{i,s}$ are the $s$-th components of $\tilde{\y}_{i}$, $\x_r$, and $\tilde{\z}_{i}$. 
The result of this diagonalization is a decomposition of the MIMO Y-channel into $N$ parallel SISO Y-channels. 
Now let us describe the transmission strategies to be used over these sub-channels.

\subsection{Transmission strategies}
In this subsection, we describe the construction of $\a_i$ and $\x_r$, and the decoding strategies used by the users and the relay.

\subsubsection{Bi-directional strategy}
\label{Sec:BiDirStrategy}
Consider the 2-cycle $(i,j)$. For this cycle, users $i$ and $j$ set  $a_{i,s}=u_{ij}$ and $a_{j,s}=u_{ji}$, where $u_{ij},u_{ji}\in\mathbb{C}$ are codeword symbols. The remaining user $k$ sets $a_{k,s}=0$. The relay receives 
\begin{align*}
y_{r,s}=\alpha_iu_{ij}+\alpha_ju_{ji}+z_{r,s}
\end{align*} 
from which it computes $L_{ij}(u_{ij},u_{ji})=\alpha_iu_{ij}+\alpha_ju_{ji}$, and sets $\x_{r,s}=\gamma_sL_{ij}(u_{ij},u_{ji})$ where $\gamma_s$ is a power allocation parameter. User $i$ receives
\begin{align*}
\tilde{y}_{i,s}=\gamma_sL_{ij}(u_{ij},u_{ji})
+\tilde{z}_{i,s},
\end{align*}
from which $u_{ji}$ is decoded after self-interference cancellation. User $j$ obtains his desired signals similarly. If each user wants to achieve $d$ DoF in this transmission, then a bundle of $d$ sub-channels is used in both the uplink and downlink.

\subsubsection{Cyclic strategy}
\label{Sec:kCycStrategy}
Consider the 3-cycle $(i,j,k)$. In this case, users $i$, $j$, and $k$ use 2 sub-channels $s_1$ and $s_2$, and set $(a_{i,s_1},a_{i,s_2})=(v_{ij},0)$, $(a_{j,s_1},a_{j,s_2})=(v_{jk},v_{jk})$, and $(a_{k,s_1},a_{k,s_2})=(0,v_{ki})$, respectively. The relay receives the following signals
\begin{align*}
y_{r,s_1}&=\alpha_{i}v_{ij}+\alpha_{j}v_{jk}+z_{r,s_1},\\
y_{r,s_2}&=\alpha_{j}v_{jk}+\alpha_{k}v_{ki}+z_{r,s_2}.
\end{align*}
It computes the sums $L_{ij}(v_{ij},v_{jk})=\alpha_{i}v_{ij}+\alpha_{j}v_{jk}$ and $L_{jk}(v_{jk},v_{ki})=\alpha_{j}v_{jk}+\alpha_{k}v_{ki}$, and sends $\x_{r,s_1}=\gamma_{s_1}L_{ij}(v_{ij},v_{jk})$ and $\x_{r,s_2}=\gamma_{s_2}L_{jk}(v_{jk},v_{ki})$. User $i$ receives 
\begin{align*}
\tilde{y}_{i,s_1}
&=\gamma_{s_1}L_{ij}(v_{ij},v_{jk})+\tilde{z}_{i,s_1},\\
\tilde{y}_{i,s_1}
&=\gamma_{s_2}L_{jk}(v_{jk},v_{ki})+\tilde{z}_{i,s_2}.
\end{align*} 
Then, users $i$ can extract both unknown signals. Namely, user $i$ decodes $v_{jk}$ from $y_{i,s_1}$ after subtracting self-interference, and then decodes $v_{ki}$ after subtracting $v_{jk}$ which has already been decoded. If each user wants to send $d$ streams to the next users in the cycle, then a bundle of $2d$ sub-channels is used for each signal-pair in the uplink and in the downlink.

\subsubsection{Uni-directional strategy}
The uni-directional strategy is a simple decode-forward strategy (or amplify-forward strategy \cite{ParkAlouiniParkKo}). In this strategy, each user sends $d$ symbols to the desired destination over $d$ sub-channels in the uplink and $d$ sub-channels in the downlink. 

\begin{table}
\centering
\begin{tabular}{| c || c | c |c |}
    \hline
    Transmission & Dimensions & Symbols & Efficiency\\
    strategy & required & delivered & (DoF/dimension)\\\hline
    Bi-directional & $1$ & $2$ & $2$\\\hline
    Cyclic & $2$ & $3$ & $3/2$\\\hline
    Uni-directional & $1$ & $1$ & $1$\\\hline
    \end{tabular}
    \caption{The proposed transmission strategies for the MIMO Y-channel listed in decreasing order of efficiency.}
    \label{Tab:Schemes}
\vspace{-.6cm}
\end{table}

These strategies along with their corresponding efficiencies are collected in Table \ref{Tab:Schemes}. The next goal is to distribute the $N$ sub-channels optimally between the users. This problem can be interpreted as a resource allocation problem where the available resources are the $N$ sub-channels. An optimal resource allocation strategy is provided in the next subsection.

\subsection{Resource allocation and transmission}
We need to develop a resource allocation strategy which guarantees the achievability of any DoF tuple $\d\in\mathcal{D}$ \eqref{Eq:DoFBoundThm}. Recall that the outer bound $\mathcal{D}$ is described by DoF constraints that do not constitute any cycles. On the other hand, a DoF tuple $\d\in\mathcal{D}$ might constitute cycles. Thus, the optimal resource allocation strategy should resolve such cycles.  

Let us consider any DoF tuple $\d\in\mathcal{D}$, which we need to achieve using the strategies listed in Table \ref{Tab:Schemes}. When distributing the sub-channels between the transmission strategies, we take into account their efficiency. Therefore, we start with the bi-directional strategy, followed by the cyclic strategy, and finally we finish with the uni-directional strategy.

\subsubsection{Resource allocation for the bi-directional strategy}
\label{Sec:ResAllocBiDir}
For each 2-cycle $(i,j)$, $i<j$, we allocate the DoF to the bi-directional strategy according to
\begin{align}
\label{eq:DoF2Cycle}
d_{(i,j)}=d_{(j,i)}=\min\left\{d_{ij},d_{ji}\right\}.
\end{align}
Using this allocation, bi-directional communication over cycle $(i,j)$ requires $d_{(i,j)}$ sub-channels. The involved users in this cycle ($i$ and $j$) perform bi-directional communication via the relay over $d_{(i,j)}$ sub-channels as described in Section \ref{Sec:BiDirStrategy}.

\subsubsection{Resource allocation for the cyclic strategy}
\label{Sec:ResAllocCyc}
After allocating resources to 2-cycles, 3 components of $\d$ are achieved. The residual DoF tuple $\d'$ with components $d_{ij}'=d_{ij}-d_{(i,j)}$ might constitute a 3-cycle. Consider a 3-cycle $(i,j,k)$. We allocate the DoF to the cyclic strategy corresponding to this 3-cycle as follows
\begin{align}
\label{eq:dc3m}
d_{(i,j,k)}=d_{(k,i,j)}=d_{(j,k,i)}=\min\left\{d_{ij}',d_{jk}',d_{ki}'\right\}.
\end{align}
Using this allocation, communication over the 3-cycle $(i,j,k)$ requires $2d_{(i,j,k)}$ sub-channels, and the transmission of the corresponding signals is done as described in Section \ref{Sec:kCycStrategy}.

\subsubsection{Resource allocation for the uni-directional strategy}
After considering all cycles of length 2 and 3, there might still remain some residual DoF tuple that need to be achieved. This is achieved using the uni-directional strategy. The remaining DoF to be achieved by the uni-directional strategy from user $i$ to user $j$ can be expressed as
\begin{align}
\label{eq:du}
d_{ij}^u=d_{ij}-d_{(i,j)}-d_{(i,j,k)}.
\end{align}
At this point, the description of the resource allocation strategy is complete. Next, we show that this strategy achieves any DoF tuple $\d$ in the DoF region $\mathcal{D}$ defined in Theorem \ref{Thm:DoF}.

\subsection{Optimality}
Let us start by writing the required number of sub-channels by this resource allocation strategy. Let $\mathcal{S}_2$ and $\mathcal{S}_3$ denote the set of 2-cycles and 3-cycles given by $\{(1,2),(1,3),(2,3)\}$ and $\{(1,2,3),(1,3,2)\}$, respectively. Since bi-directional communication over a cycle $(i,j)$ requires $d_{(i,j)}$ sub-channels (cf. Section \ref{Sec:ResAllocBiDir}), then the number of sub-channels required for all 2-cycles is $\sum_{\c\in\mathcal{S}_2}d_{\c}$. Similarly, the cyclic strategy  corresponding to the 3-cycle $(i,j,k)$ requires $2d_{(i,j,k)}$ sub-channels (cf. Section \ref{Sec:ResAllocCyc}), and hence the number of sub-channels required for all 3-cycles is $2\sum_{\c\in\mathcal{S}_3}d_{\c}$. The number of sub-channels required by the uni-directional strategy  is the sum over all $i\neq j$ of $d_{ij}^u$, 
which can be written as 
$$\sum_{i\in\mathcal{K}}\sum_{j\in\mathcal{K}\setminus\{i\}} d_{ij}^u=\sum_{i\in\mathcal{K}}\sum_{j\in\mathcal{K}\setminus\{i\}} d_{ij}-2\sum_{\c\in\mathcal{S}_2}d_{\c}
-3\sum_{\c\in\mathcal{S}_3}d_{\c}$$ by \eqref{eq:du}. 
By adding, we can express the total number of required sub-channels as
\begin{align}
N_s
\label{eq:Ns2}
&=\sum_{i\in\mathcal{K}}\sum_{j\in\mathcal{K}\setminus\{i\}}d_{ij}-\sum_{\c\in\mathcal{S}_2}d_{\c}
-\sum_{\c\in\mathcal{S}_3}d_{\c}.
\end{align}
This is the required number of sub-channels for our strategy. To be able to implement this scheme in a MIMO Y-channel with $N$ sub-channels, we need the condition $N_s\leq N$ to hold for any $\d\in\mathcal{D}$. To show that $N_s\leq N$ for any $\d\in\mathcal{D}$, we need to show that \eqref{eq:Ns2} satisfies properties (a) and (b). Note that the DoF of all cycles appear in \eqref{eq:Ns2} with a negative sign. This is sufficient to resolve all cycles. First, all 2-cycles $(i,j)$ are resolved by $-d_{(i,j)}$, and the result after subtracting the DoF of these 2-cycles is (cf. \eqref{eq:DoF2Cycle})
\begin{align}
\label{NsNo2Cycles}
N_s&=\sum_{i=1}^2\sum_{j=i+1}^3\max\{d_{ij},d_{ji}\}
-\sum_{\c\in\mathcal{S}_3}d_{\c}.
\end{align}
Clearly, \eqref{NsNo2Cycles} does not have 2-cycles. However, it might have 3-cycles. Suppose that it has the 3-cycle $(1,2,3)$, i.e., the first sum leads to $d_{12}+d_{23}+d_{31}$. This 3-cycle is resolved by the term $-d_{(1,2,3)}$. Assuming $d_{(1,2,3)}=d_{12}-d_{(1,2)}$ (cf. \eqref{eq:dc3m}, other two cases follow similarly) and substituting in $N_s$ yields
\begin{align}
N_s&=d_{12}+d_{23}+d_{31}-d_{(1,2,3)}-d_{(1,3,2)}\\
\label{eq:N_s3CycleResolve}
&=d_{12}+d_{23}+d_{31}-d_{12}+d_{(1,2)}-d_{(1,3,2)}.
\end{align}
Since we have the 3-cycle $(1,2,3)$, this implies that $d_{12}\geq d_{21}$ and hence $d_{(1,2)}=d_{21}$. Substituting in \eqref{eq:N_s3CycleResolve}, we get
\begin{align}
N_s&=d_{23}+d_{31}+d_{21}-d_{(1,3,2)}.
\end{align}
As a result, the term $-d_{(1,2,3)}$ resolves the 3-cycle $(1,2,3)$ by replacing $d_{12}+d_{23}+d_{31}$ by $d_{21}+d_{23}+d_{31}$ which does not constitute a 3-cycle. Similarly, the term $-d_{(1,3,2)}$ resolves the cycle $(1,3,2)$. If one of these 3-cycles do not exist for the given $\d\in\mathcal{D}$, the corresponding DoF is zero by \eqref{eq:dc3m}. After taking into account both 3-cycles, $N_s$ becomes of the form
\begin{align}
\label{eq:Ns4Cycle}
N_s&=d_{ij}+d_{ik}+d_{jk},
\end{align}
for distinct $i,j,k\in\mathcal{K}$. This $N_s$ is the sum of 3 components of $\d\in\mathcal{D}$ that constitute no cycles. Thus, $N_s$ satisfies properties (a) and (b). By \eqref{Eq:DoFBoundThm}, this $N_s$ has to be less than $N$ for any $\d\in\mathcal{D}$. Thus, any $\d\in\mathcal{D}$ is achievable\footnote{In this analysis, we have assumed that $\d$ has integer-valued components. DoF tuples $\d\in\mathcal{D}$ with non-integer-valued components can be achieved by considering channel extension in time as in \cite{AvestimehrKhajehnejadSezginHassibi}.}. This concludes the proof of achievability of Theorem \ref{Thm:DoF}.

\section{Sub-optimality of channel separation}
\label{Sec:SubChannelSep}
The optimal scheme for the Y-channel requires the use of the cyclic strategy (communication over 3-cycles), which in turn requires coding jointly over 2 sub-channels. Thus, the sub-channels have to be considered jointly. If we use a channel separation approach instead, where the signals transmitted over a sub-channels can be decoded by only observing this particular sub-channel, then the cyclic strategy has to be avoided. This separation approach turns out to be sub-optimal. We have seen in Section \ref{Sec:BiDirStra3User} that using the bi-directional and uni-directional strategies (which do not require joint encoding over multiple sub-channels) is not sufficient to achieve the DoF region.

However, a channel separation approach is optimal in terms of {\it sum-DoF}. If we are not interested in the DoF trade-off between different DoF component, but we are rather interested in the sum-DoF, then the bi-directional strategy suffices. To show this, note that the DoF region $\mathcal{D}$ in \eqref{Eq:DoFBoundThm} implies that the sum-DoF is given by $d_\Sigma=2N$. This can be shown by summing up the bounds corresponding to $\p=(1,2,3)$ and $\p=(3,2,1)$ in Theorem \ref{Thm:DoF}. To achieve $2N$ DoF in total, the resources ($N$ sub-channels) can be distributed among the $2$-cycles of the Y-channel in any desired manner. Then, each pair of users in a 2-cycle use the bi-directional strategy to exchange two signals (one signal in each direction) over each sub-channel assigned to this 2-cycle. We have $N$ sub-channels in total, and thus, this strategy achieves $2N$ DoF.

\section{Conclusion}
\label{Sec:Conclusion}
We have characterized the DoF region of the MIMO Y-channel with $N$ antennas at the relay and $M\geq N$ antennas at the users. The DoF region is proved to be achievable by using channel diagonalization in addition to a combination of bi-directional, cyclic, and uni-directional communication strategies. The bi-directional and cyclic strategies use compute-forward at the relay (physical-layer network-coding), while the uni-directional strategy is based on decode-forward. This combination of strategies is optimized by using a simple resource allocation approach. The resulting optimized scheme achieves the DoF region of the channel. As a by-product, we conclude that the MIMO Y-channel is inseparable. Thus, in general, one has to code over multiple sub-channels to achieve the optimal performance. The results of this work apply for the $K$-user case, and will be presented in a longer journal version of this paper due to lack of space. Note that the DoF region of the case $M< N$ has not been characterized to-date, and is an interesting problem for future work.

\bibliography{myBib}

\begin{thebibliography}{10}
\providecommand{\url}[1]{#1}
\csname url@samestyle\endcsname
\providecommand{\newblock}{\relax}
\providecommand{\bibinfo}[2]{#2}
\providecommand{\BIBentrySTDinterwordspacing}{\spaceskip=0pt\relax}
\providecommand{\BIBentryALTinterwordstretchfactor}{4}
\providecommand{\BIBentryALTinterwordspacing}{\spaceskip=\fontdimen2\font plus
\BIBentryALTinterwordstretchfactor\fontdimen3\font minus
  \fontdimen4\font\relax}
\providecommand{\BIBforeignlanguage}[2]{{%
\expandafter\ifx\csname l@#1\endcsname\relax
\typeout{** WARNING: IEEEtran.bst: No hyphenation pattern has been}%
\typeout{** loaded for the language `#1'. Using the pattern for}%
\typeout{** the default language instead.}%
\else
\language=\csname l@#1\endcsname
\fi
#2}}
\providecommand{\BIBdecl}{\relax}
\BIBdecl

\bibitem{Evans_IoT}
D.~Evans, ``{The Internet of Things: How the next evolution of the internet is
  changing everything},'' in \emph{Cisco Internet Business Solutions Group
  (IBSG) technical report}, April 2011.

\bibitem{KimDevroyeMitranTarokh}
S.~Kim, N.~Devroye, P.~Mitran, and V.~Tarokh, ``{Comparisons of bi-directional
  relaying protocols},'' in \emph{Proc. of the IEEE Sarnoff Symposium},
  Princeton, NJ, Apr. 2008.

\bibitem{AvestimehrSezginTse}
A.~S. Avestimehr, A.~Sezgin, and D.~Tse, ``{Capacity of the two-way relay
  channel within a constant gap},'' \emph{European Trans. in
  Telecommunications}, vol.~21, no.~4, pp. 363--374, 2010.

\bibitem{ShaqfehZafarAlnuweiriAlouini}
M.~Shaqfeh, A.~Zafar, H.~Alnuweiri, and M.-S. Alouini, ``{Joint opportunistic
  scheduling and network coding for bidirectional relay channel},'' in
  \emph{Proc. of IEEE International Symposium on Info. Theory (ISIT)},
  Istanbul, Turkey, 2013.

\bibitem{ChaabanSezginAvestimehr_YC_SC}
A.~Chaaban, A.~Sezgin, and A.~S. Avestimehr, ``{Approximate sum capacity of the
  Y-channel},'' \emph{IEEE Trans. on Info. Theory}, vol.~59, no.~9, pp.
  5723--5740, Sept. 2013.

\bibitem{MatthiesenZapponeJorswieck}
B.~Matthiesen, A.~Zappone, and E.~A. Jorswieck, ``{Spectral and energy
  efficiency in 3-way relay channels with circular message exchanges},'' in
  \emph{Proc. of 11th Internation Symposium on Wireless Communication Systems
  (ISWCS)}, Barcelona, Spain, 2014.

\bibitem{GunduzYenerGoldsmithPoor_IT}
D.~G\"und\"uz, A.~Yener, A.~Goldsmith, and H.~V. Poor, ``{The multi-way relay
  channel},'' \emph{IEEE Trans. on Info. Theory}, vol.~59, no.~1, pp. 51--63,
  Jan. 2013.

\bibitem{MokhtarMohassebNafieElGamal}
M.~Mokhtar, Y.~Mohasseb, M.~Nafie, and H.~El-Gamal, ``{On the deterministic
  multicast capacity of bidirectional relay networks},'' in \emph{Proc. of the
  2010 IEEE Info. Theory Workshop (ITW)}, Dublin, Aug. 2010.

\bibitem{SezginAvestimehrKhajehnejadHassibi}
A.~Sezgin, A.~S. Avestimehr, M.~A. Khajehnejad, and B.~Hassibi,
  ``{Divide-and-conquer: Approaching the capacity of the two-pair bidirectional
  Gaussian relay network},'' \emph{IEEE Trans. on Info. Theory}, vol.~58,
  no.~4, pp. 2434--2454, Apr. 2012.

\bibitem{OngKellettJohnson_IT}
L.~Ong, C.~M. Kellett, and S.~J. Johnson, ``{On the equal-rate capacity of the
  AWGN multiway relay channel},'' \emph{IEEE Trans. on Info. Theory}, vol.~58,
  no.~9, pp. 5761--5769, Sept. 2012.

\bibitem{CoverThomas}
T.~Cover and J.~Thomas, \emph{{Elements of information theory (Second
  Edition)}}.\hskip 1em plus 0.5em minus 0.4em\relax John Wiley and Sons, Inc.,
  2006.

\bibitem{ChaabanSezgin_ISIT12_Y}
A.~Chaaban and A.~Sezgin, ``{Signal space alignment for the Gaussian
  Y-channel},'' in \emph{Proc. of IEEE International Symposium on Info. Theory
  (ISIT)}, Cambridge, MA, July. 2012, pp. 2087--2091.

\bibitem{ZewailNafieMohassebGamal}
A.~A. Zewail, M.~Nafie, Y.~Mohasseb, and H.~El-Gamal, ``{Achievable degrees of
  freedom region of MIMO relay networks using detour schemes},'' in \emph{Proc.
  of IEEE International Conference on Communications (ICC)}, Sydney, Australia,
  2014.

\bibitem{LeeLimChun}
N.~Lee, J.-B. Lim, and J.~Chun, ``{Degrees of freedom of the MIMO Y channel:
  Signal space alignment for network coding},'' \emph{IEEE Trans. on Info.
  Theory}, vol.~56, no.~7, pp. 3332--3342, Jul. 2010.

\bibitem{ChaabanOchsSezgin}
A.~Chaaban, K.~Ochs, and A.~Sezgin, ``{The degrees of freedom of the MIMO
  Y-channel},'' in \emph{Proc. of IEEE International Symposium on Info. Theory
  (ISIT)}, Istanbul, July 2013.

\bibitem{NazerGastpar}
B.~Nazer and M.~Gastpar, ``{Compute-and-forward: Harnessing interference
  through structured codes},'' \emph{IEEE Trans. on Info. Theory}, vol.~57,
  no.~10, pp. 6463--6486, Oct. 2011.

\bibitem{ChaabanOchsSezgin_EW}
A.~Chaaban, K.~Ochs, and A.~Sezgin, ``{Simultaneous diagonalization: On the DoF
  region of the K-user MIMO multi-way relay channel},'' in \emph{European
  Wireless 2014}, Barcelona, Spain, May 2014.

\bibitem{CadambeJafar_Inseperability}
V.~Cadambe and S.~A. Jafar, ``{Parallel Gaussian interference channels are not
  always separable},'' \emph{IEEE Trans. on Info. Theory}, vol.~55, no.~9, pp.
  3983--3990, Sep. 2009.

\bibitem{CadambeJafar_KUserIC}
V.~R. Cadambe and S.~A. Jafar, ``{Interference alignment and the degrees of
  freedom for the K user interference channel},'' \emph{IEEE Trans. on Info.
  Theory}, vol.~54, no.~8, pp. 3425--3441, Aug. 2008.

\bibitem{CoverElgamal}
T.~M. Cover and A.~El-Gamal, ``{Capacity theorems for the relay channel},''
  \emph{IEEE Trans. on Info. Theory}, vol. IT-25, no.~5, pp. 572--584, Sep.
  1979.

\bibitem{ShaqfehQahtaniAlnuweiri}
M.~Shaqfeh, F.~Al-Qahtani, and H.~Alnuweiri, ``{Optimal relay selection for
  decode-and-forward opportunistic relaying},'' in \emph{International
  Conference on Communications, Signal Processing, and their Applications
  (ICCSPA)}, Sharjah, UAE, Feb. 2013.

\bibitem{NamChungLee}
W.~Nam, S.-Y. Chung, and Y.~H. Lee, ``{Capacity bounds for two-way relay
  channels},'' in \emph{Proc. of the IEEE International Zurich Seminar},
  Zurich, Mar. 2008, pp. 144--147.

\bibitem{ParkAlouiniParkKo}
K.-H. Park, M.-S. Alouini, S.-H. Park, and Y.-C. Ko, ``{On the achievable
  degrees of freedom of alternate MIMO relaying with multiple AF relays},'' in
  \emph{Third International Conference on Communications and Networking
  (ComNet)}, Hammamet, Tunesia, March 2012.

\bibitem{AvestimehrKhajehnejadSezginHassibi}
A.~S. Avestimehr, M.~A. Khajehnejad, A.~Sezgin, and B.~Hassibi, ``{Capacity
  region of the deterministic multi-pair bi-directional relay network},'' in
  \emph{Proc. of the ITW}, Volos, Greece, Jun. 2009.

\end{thebibliography}

\end{document}